\documentclass[smallextended]{svjour3}       
\smartqed  
\usepackage{graphicx}
\usepackage{psfrag}
\usepackage[hypertex]{hyperref}
\usepackage{ifthen,amsmath,amssymb, bm}
\usepackage{mathptmx}      
\journalname{General Relativity and Gravitation}
\newcommand{\figwidth}{\columnwidth}

\renewcommand{\vec}[1]{\mathbf{#1}}
\newcommand{\mat}[1]{\bm{#1}}

\usepackage{color}

\newcommand{\halofit}{\textsc{halofit}\ }
\newcommand{\camb}{\textsc{camb}\ }

\def\eprinttmp@#1arXiv:#2 [#3]#4@{
\ifthenelse{\equal{#3}{x}}{\href{http://arxiv.org/abs/#1}{#1}}{\href{http://arxiv.org/abs/#2}{arXiv:#2} [#3]}}

\newcommand{\eprint}[1]{\eprinttmp@#1arXiv: [x]@}

\newcommand{\adsurl}[1]{\href{#1}{ADS}}

\newcommand{\clo}{\mathcal{O}}
\newcommand{\Mpc}{\text{Mpc}}

\newcommand{\mN}{\bm{N}}

\newcommand{\vF}{\vec{F}}

\newcommand{\vL}{\vec{L}}

\newcommand{\vk}{\vec{k}}
\newcommand{\vl}{\vec{l}}

\newcommand{\vx}{\vec{x}}
\newcommand{\vy}{\vec{y}}
\newcommand{\vz}{\vec{z}}

\newcommand{\begm}{\begin{pmatrix}}
\newcommand{\enm}{\end{pmatrix}}

\newcommand{\degree}{{}^{\rm o}}

\newcommand{\grad}{\nabla}

\newcommand{\be}{\begin{equation}}
\newcommand{\ee}{\end{equation}}

\newcommand{\ba}{\begin{eqnarray}}
\newcommand{\ea}{\end{eqnarray}}
\newcommand{\nn}{\nonumber}

\newcommand{\tr}{\text{Tr}}
\newcommand{\lp}{{\cal L}}

\newcommand{\f}{{\cal F}}           

\newcommand{\eal}{{\cal A}}         

\newcommand{\lop}{\bm{\Lambda}}          

\newcommand{\mcov}{\mat{C}}         

\newcommand{\pot}{\psi}               

\newcommand{\defl}{{\alpha}}           
\newcommand{\vdefl}{\vec{\alpha}}           

\newcommand{\sr}{\xi}               

\newcommand{\gp}{\Psi}              

\newcommand{\au}{{\Theta}}    
\newcommand{\al}{\tilde{\Theta}}            
\newcommand{\ao}{\hat{\Theta}}      

\newcommand{\vau}{\bm{\Theta}}    
\newcommand{\val}{\tilde{\vau}}            
\newcommand{\vao}{\hat{\vau}}      

\newcommand{\tl}{\tilde{T}}                 
\newcommand{\tu}{{T}}         
\newcommand{\Ql}{\tilde{Q}}                 
\newcommand{\Qu}{{Q}}         
\newcommand{\Ul}{\tilde{U}}                 
\newcommand{\Uu}{{U}}         

\newcommand{\el}{\tilde{E}}                 
\newcommand{\eu}{{E}}         
\newcommand{\bl}{\tilde{B}}                 
\newcommand{\bu}{{B}}         
\newcommand{\nse}{n}                
\newcommand{\vnse}{\vec{\nse}}

\newcommand{\hq}{{\cal H}}          
\newcommand{\vhq}{\vec{\hq}}

\newcommand{\mlensedcov}{\tilde{\mcov}^{\al\al}}
\newcommand{\munlensedcov}{{\mcov}^{\au\au}}
\newcommand{\mobscov}{\hat{\mcov}^{\ao \ao}}
\newcommand{\mnoisecov}{\mN}
\newcommand{\reconvar}{\vdefl}
\newcommand{\hreconvar}{\hat{\vdefl}}       
\newcommand{\reconvartext}{deflection angle}

\newcommand{\arcmin}{~{\rm arcmin}}

\begin{document}

\title{Weak lensing of the CMB
}


\author{Duncan Hanson \and Anthony Challinor \and Antony Lewis}


\institute{D. Hanson \and A. Challinor \and A. Lewis \at
              Institute of Astronomy and Kavli Institute for Cosmology Cambridge, Madingley Road, Cambridge, CB3 0HA, UK \\
\email{dhanson@ast.cam.ac.uk}
\and A. Challinor \at
DAMTP, Centre for Mathematical Sciences, Wilberforce Road, Cambridge CB3 OWA, UK
}

\date{Received: date / Accepted: date}

\maketitle

\begin{abstract}
The cosmic microwave background (CMB) represents a unique source for the
study of gravitational lensing.
It is extended across the entire sky, partially polarized, located
at the extreme distance of $z=1100$, and is thought to have the simple,
underlying statistics of a Gaussian random field.
Here we review the weak lensing of the CMB, highlighting
the aspects which differentiate it from the weak lensing
of other sources, such as galaxies.
We discuss the statistics of the lensing deflection field which remaps the CMB,
and the corresponding effect on the power spectra. We then focus on
methods for reconstructing the lensing deflections, describing efficient quadratic
maximum-likelihood estimators and delensing. We end by reviewing recent detections
and observational prospects.
\keywords{Cosmic Microwave Background; Gravitational Lensing}
\end{abstract}


\section{Introduction}
\label{sec:introduction}
\subsection{\textit{What is the CMB?}}
The cosmic microwave background (CMB) is a blackbody radiation
which permeates the Universe. It was released approximately
$400,000$ years after the Big Bang, when the primordial plasma
had cooled enough that neutral hydrogen atoms could form,
and the Universe became effectively transparent
to radiation.
To a very good degree of approximation, the CMB temperature
is uniform across the sky, with a value today of
$2.725\,\mathrm{K}$~\cite{2002ApJ...581..817F}.
At the level of one part in $10^{5}$, however, the CMB temperature has small
anisotropies which
are due to fluctuations of density and bulk velocity in the primordial plasma.
The measurement of these anisotropies is clearly a difficult task,
but a number of very successful experiments have driven
rapid progress in this field over the past two decades.
Modern temperature experiments such as
ACT\footnote{\url{http://www.physics.princeton.edu/act/}},
SPT\footnote{\url{http://pole.uchicago.edu/}},
and
Planck\footnote{\url{http://www.rssd.esa.int/index.php?project=planck}}
are now pushing to robust arcminute-scale measurements, while polarization experiments
like
Spider\footnote{\url{http://www.astro.caltech.edu/~lgg/spider_front.htm}},
QUIET\footnote{\url{http://quiet.uchicago.edu/}},
BICEP2
and
Planck\footnote{\url{http://www.rssd.esa.int/index.php?project=planck}}
will soon obtain low-noise maps of the
CMB polarization.
Gravitational lensing effects should become readily apparent in both types of observation,
requiring detailed modelling to interpret the data correctly, but also offering additional
information about the Universe at intermediate redshifts.

\subsection{\textit{What is special about the CMB in the context of lensing?}}
\begin{figure}
\begin{center}
\psfrag{tT}{$\tilde{T}$}
\psfrag{tU}{$\tilde{T} - T\ (\!\times 5)$}
\psfrag{thtdg}{$\theta\ (\rm deg.)$}
\psfrag{psi}{$\psi$}
\includegraphics[width=\figwidth]{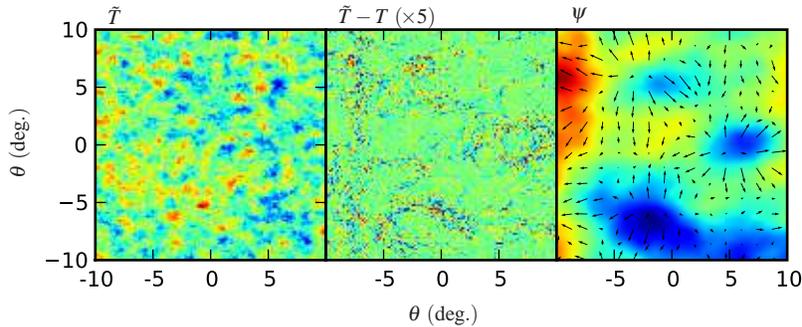}
\caption{Left: lensed CMB realization. Middle: difference map between lensed and unlensed CMB. Right: realization of $\psi$ for the lensing, and an overlay of its gradient, the deflection angle.}
\label{fig:lensed_unlensed_map}
\end{center}
\end{figure}

In the context of lensing, the CMB
can be thought of as a single source ``plane'' with several unique features:
\begin{enumerate}
\item{} It is extended across the entire sky.
\item{} It probes the matter distribution at relatively higher redshifts
($z \sim 2$) than can be studied with e.g. galaxy lensing: the last-scattering
surface from which the CMB was released ($z\sim 1100$) is
the most distant light source we will ever be able to see.
\item{} It is partially linearly polarized, which provides two additional lensed observables.
The local quadrupole of the radiation field at the last scattering
surface generates linear polarization through the mechanism of
Thomson scattering, at a level of $10\%$ of the temperature
anisotropies \cite{Hu:1997hv}.
The observable CMB is therefore characterized by three numbers as a function of position
on the sky -- the total radiation intensity, and two Stokes parameters
describing the amplitude and direction of the linear polarization.
\item{ To a good degree of approximation the source fluctuations
may be treated as those of a Gaussian random field. The signal
along any particular direction is essentially random, but the correlations
between directions at different angular separations is encoded in the power spectrum, which completely characterizes the signal.
A typical CMB temperature realization, for example, is plotted in Fig.~\ref{fig:lensed_unlensed_map}}. \end{enumerate}

\subsection{\textit{Why study CMB lensing?}}
To someone interested in the primary CMB,
the lensing of the CMB is important
as a contaminant. As we will discuss in
Sec.~\ref{sec:lensing_effects_on_cmb_observables}, lensing has a small
but non-negligible effect on the CMB power spectra, and must be
accounted for in modern parameter analyses to obtain unbiased constraints.
Perhaps more importantly, lensing effects generate a curl-like ($B$ mode)
polarization pattern on the sky which acts as a limiting source of
confusion for low-noise polarization experiments targeting the signal from primordial gravitational waves~\cite{Knox:2002pe,Seljak:2003pn}.
This confusion can be reduced with
an accurate cleaning of the lensing-induced signal,
which we will discuss in Sec.~\ref{sec:lens_reconstruction}.

Apart from being a nuisance for traditional observables,
the lensing of the CMB can act as an additional source of information.
A typical analysis of the CMB assumes Gaussianity and statistical isotropy,
in which case the power spectrum is the only quantity of interest.
As we shall discuss, lensing can be thought of as introducing into the
CMB small amounts of non-Gaussianity (when marginalized over realizations
of the lenses) or statistical anisotropy (for a fixed distribution of lenses).
This effectively
\textit{introduces} information into the CMB, contained in the higher-order statistics (for the non-Gaussian viewpoint) and the off-diagonal elements
of its covariance matrix (for the anisotropy viewpoint).
With only one CMB sky to observe and interpret, both these viewpoints are
useful.
The additional information from lensing probes the state of the Universe at intermediate
redshifts ($z \sim 2$). This can be used to break parameter degeneracies and place improved
constraints on quantities that affect the geometry or density perturbations at late times, such as the dark energy equation of state and portion of the energy budget in massive neutrinos.
An optimal analysis of
lensing effects with the data from the Planck satellite, for example, will enable us to measure the
sum of neutrino masses to $\sim 0.1\, {\rm eV}$, while lens reconstruction with
a next-generation polarization mission such as EPIC/CMBPol can constrain
the sum to $0.05\, {\rm eV}$ or better~\cite{Kaplinghat:2003bh,Lesgourgues:2005yv,dePutter:2009kn}. This is
an interesting limit, close to the minimum value for the sum of the masses
suggested
by terrestrial oscillation measurements in the normal hierarchy.

\section{Observables}
\label{sec:notation}
Here we will briefly review the mathematical
description of CMB observables and establish our notation.
Throughout this review
we will use an overtilde to represent lensed quantities
For quantitative calculations we assume that the Universe is described
by a flat, concordance $\Lambda$CDM cosmology with parameters given
by the WMAP 5-year best-fit model.

The CMB radiation pattern on the sky is conveniently expressed in terms
of the Stokes parameters $T$, $Q$ and $U$, which are measures of the total
intensity ($T$; here expressed as an equivalent blackbody temperature
relative to the background, uniform CMB) and linear polarization ($Q$ and $U$).
Rotation of the measurement coordinates does not affect $T$,
however the $Q$ and $U$ Stokes parameters mix according to
\be
[Q \pm iU]_{\sr} = e^{\pm 2 i \sr} [Q \pm iU]_0 \, ,
\ee
where $\sr$ is the angle of the rotation\footnote{This is for Stokes
parameters defined on a right-handed $x$-$y$ basis with $z$ along the propagation
direction, and for a left-handed
rotation about $z$.}.
Thus, $Q \pm iU$ is a spin\ $\pm 2$ quantity.
A rotation of $45$ degrees converts $Q \rightarrow U$,
and a $90$-degree rotation results in a sign flip.
For more information on this notation, see for example~\cite{Partridge3K1995}.

The statistics of the CMB are often simplest to study as
a function of scale rather than angular position.
On a patch of the sky which is small enough that it may be
treated as flat, the Fourier modes provide an appropriate basis.
In this review, we will take the convention that
\ba
\tu(\vl) &=& \int \frac{d^2 \vx}{2\pi} \tu(\vx) e^{-i \vl \cdot \vx} \nn \\
\left[ \eu(\vl) \pm i\bu(\vl) \right]  &=& -\int \frac{d^2 \vx}{2\pi} [\Qu \pm i\Uu] e^{\mp 2i \sr_{\vl}} e^{-i \vl \cdot \vx}.
\ea
Here $\sr_{\vl}$ is the angle between $\vl$ and the $x$-axis.
The placement of the $e^{\mp 2i \sr_{\vl}}$ factor in the
polarization transform rotates the Fourier transforms of the Stokes
parameters to a basis adapted to the Fourier wavevector,
resulting in $E$ and $B$ fields which
are spin-$0$. This makes $E$ and $B$ independent of any measurement conventions,
and so they are the natural basis to perform calculations in. The $E$/$B$-mode decomposition
is also useful for studying the origin of fluctuations in the early Universe:
$B$-modes are not generated by scalar perturbations, and so the detection of non-zero
primordial $B$-mode power could in principle be a ``smoking gun'' for primordial tensor
modes \cite{Seljak:1996gy,Kamionkowski:1996zd}.

The primary, unlensed CMB fluctuations are thought to be well approximated
as a statistically-isotropic Gaussian random field with zero mean.
This means that in the Fourier domain,
each mode is drawn from a (complex) Gaussian, with variance which depends
only on the length of the wavevector, and not on its direction. The
length-dependent part is known as the power spectrum $C_l$. In the Fourier basis, for example
we have
\be
\langle X(\vl) Z^{*}(\vl') \rangle = \delta(\vl - \vl') C_{l}^{XZ},
\ee
where $X$, $Z$ can be $T$, $E$ or $B$, and $l = |\vl|$.
If parity is respected in the mean,
$C_l^{TB}$ and $C_l^{EB}$ must vanish, but note that there is a non-zero
cross-correlation between the temperature and the $E$-mode polarization.
Because the modes are drawn from a Gaussian distribution, they
are completely characterized by their two-point statistics. Higher order
odd-point functions are zero, and the even-point functions can be computed
from the power spectra using Wick's theorem.


We now proceed to describe the large-scale structures responsible for the lensing of the CMB.

\section{Lensing potential}
\label{sec:lensing_potential}

\subsection{\textit{Lensing deflection}}
The effect of lensing on the CMB can be described in the
flat-sky limit as a remapping of the primary (unlensed)
CMB given by
\ba
\tl(\vx) &=& \tu(\vx + \vdefl(\vx)) \\
\left[\Ql \pm i\Ul \right](\vx)  &=& \left[\Qu \pm i\Uu \right](\vx + \vdefl(\vx))
\label{eqn:lens_remapping}
\ea
where $\vdefl(\vx)$ is a field of deflection vectors.

The observed deflection vector is given by a sum of all the deflections by lenses between ourselves and the last
scattering surface; mathematically this is given by an integral over the standard weak lensing equation along the photon path. In a flat universe, and under the Born approximation (taking the integral along the unperturbed path) we have
%
\be
\vdefl(\vx) = -2 \int_0^{\chi_*} d\chi
\frac{\chi_* - \chi}{\chi_* } \nabla_{\perp}
\Psi(\chi \vx; \eta_0 - \chi ),
\label{eqn:lensing_deflection}
\ee
where $\gp$ is the Weyl potential,
$\chi_{*}$ is the conformal distance to last scattering,
and $\eta_0$ is the conformal time today.
Here, the gradient is taken transverse to the propagation direction.
In Einstein gravity with vanishing anisotropic stresses,
the Weyl potential is simply the Newtonian potential. More generally,
in any metric theory of gravity, $\gp$ is given in
terms of the Newtonian-gauge potentials
$\phi_{\mathrm{N}}$ and $\psi_{\mathrm{N}}$ by $\Psi = (\phi_{\mathrm{N}} + \psi_{\mathrm{N}})/2$: it is the scalar field which determines the
Weyl tensor (non-local part of the Riemann tensor) in the presence of scalar perturbations. For further details and the simple generalization to a non-flat Universe see~\cite{Lewis:2006fu}.

\subsection{\textit{Order of magnitude}}
It is useful to get a feel for the order of magnitude of the lensing effect on the CMB, and why it is important.
We can think of the large scale structure of the Universe as
a collection of potential wells, crudely approximated as those due to
point-source masses.
Each gives an observed deflection angle of
\be
|\vdefl^{\rm ps}(\vx)| = 4 \left( \frac{\chi_{*} - \chi}{\chi_*} \right) \Psi^{\rm ps}(\chi \vx) \, ,
\label{eqn:ps_deflection}
\ee
where $\chi$ is the conformal distance to the point source and $\gp^{\rm ps}(\chi \vx)$ is the potential
at closest approach. Thus nearby lenses have the largest contribution from
the geometric factor $(\chi_* - \chi)/\chi_*$, though
the linear potentials are actually somewhat smaller at low redshift due to the suppression of perturbation growth by the dark energy.
During matter domination the geometric factor is given by $(1-\chi / \chi_*) \sim [1/(1+z)]^{1/2}$,
which depends somewhat weakly on redshift, so we will ignore it for the rest
of this order-of-magnitude calculation.

The potentials are nearly scale-invariant on large scales,
but become small on scales below the turnover in the matter power spectrum.
We therefore need to consider lensing by perturbations from the largest observable scale
$\sim \chi_{*}\sim 14000\, \Mpc$ down to $\sim 300\, \Mpc$. The number of lenses along the
 line of sight also depends on scale: there is only one of the largest lenses,
but $\sim 14000/300\sim 50$ of the smallest ones, so the smaller lenses contribute
most to the lensing deflection.
The size of each deflection is set by the Newtonian potentials, which have variance
$\sim (2.7 \times 10^{-5})^2$ per unit log range in scale, so typical deflections
have $|\vdefl| \sim 10^{-4}$. Taking each lens to be independent, the total r.m.s.
deflection should therefore be $\sim 50^{1/2}\times 10^{-4} \sim 7\times 10^{-4}$,
or about $2\arcmin$.

The coherence scale of the deflection field is given by the angular extent
of a typical $300\,{\rm Mpc}$ structure. At a distance $\chi=7000\,\Mpc$
(halfway to the last-scattering surface) this is
$(300 / 7000) \sim 2\degree$.

\begin{figure}
\begin{center}
\includegraphics[width=2.35in]{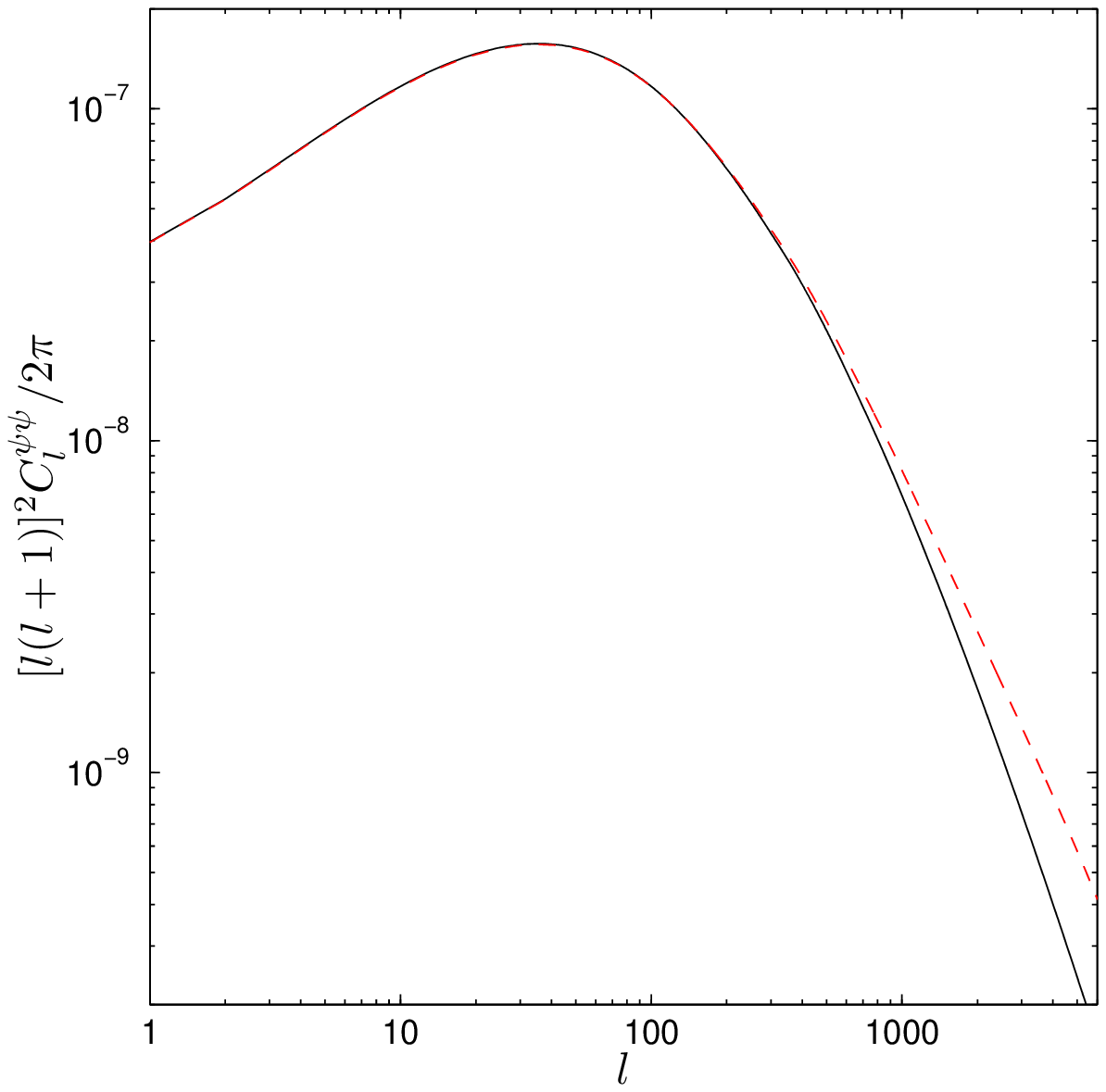}
\includegraphics[width=2.3in]{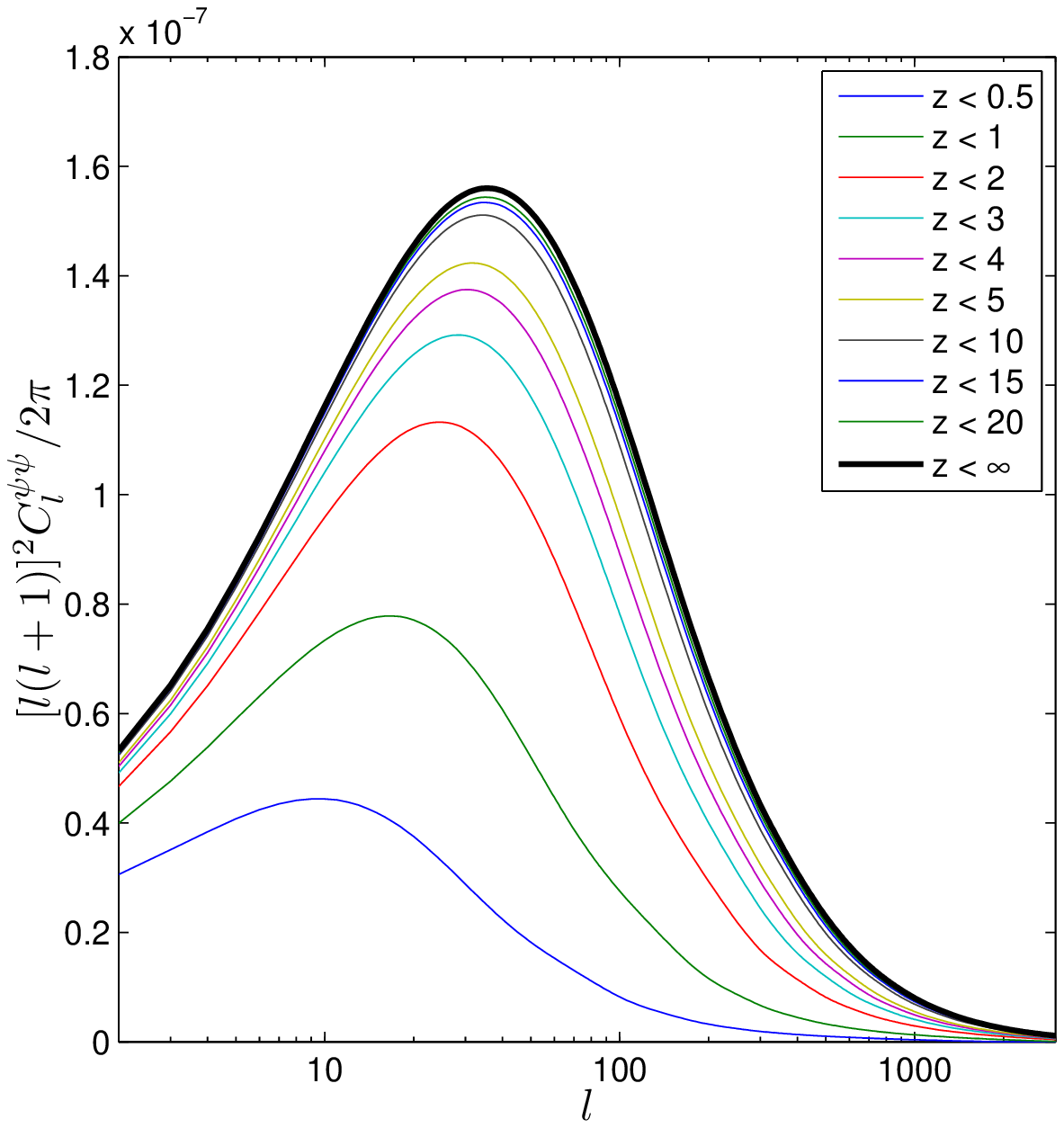}
\caption{Left: angular power spectrum of the lensing deflection angle in linear theory (solid), and with the \halofit non-linear corrections given by Eq.~\eqref{eqn:halofit_corrections} (dashed). Right: cumulative contributions to the power spectrum as a function of maximum redshift.}
\label{fig:clphiphi}
\label{fig:clphiphiz}
\end{center}
\end{figure}

These order of magnitude estimates are borne out by more accurate
calculations. The RMS deflection angle for the WMAP 5-year cosmology is
$2.7\arcmin$. The scale dependence (i.e.\ angular power spectrum) and
cumulative redshift dependence of the deflection power
are shown in Fig.~\ref{fig:clphiphiz}.
Nearby objects tend
to contribute to the lensing potential on larger angular scales.
The deflection power peaks in the region $20 < l < 100$, corresponding
to a coherence length of a few degrees.

%

\subsection{\textit{Lensing potential}}
Rather than using the deflection field directly, weak lensing studies are often cast in terms of the
%
magnification matrix $A_{ij}$ defined by
\be
A_{ij}\equiv \delta_{ij} + \frac{\partial}{\partial\theta_i} \defl_j = \begm 1-\kappa-\gamma_1\, && -\gamma_2 + \omega \\ -\gamma_2 -\omega\, && 1-\kappa+\gamma_1 \enm ,
\ee
where the quantities $\kappa$, $\gamma_1~+~i\gamma_2$, and $\omega$ are known
as the convergence, (complex) shear, and field rotation respectively.
In galaxy lensing
the magnification matrix is useful because
in many cases this can be approximated as being constant over the
angular extent of the source. Shear then produces an additional elliptical
distortion of the projected shapes of galaxies, while convergence
changes their apparent size and density on the sky.
For galaxy lensing the shear is the more
useful observable since observing the convergence would require knowledge of the
unlensed distribution of galaxies.
In CMB lensing, the degree-scale acoustic features are large compared to the
coherence size of the magnification matrix, so the latter cannot be assumed
constant over typical CMB features. Moreover,
the statistics of the source plane are
well known (the CMB is nearly a Gaussian random field), and so the convergence is just as useful as shear.
For these reasons CMB lensing studies usually work directly with the lensing
deflection field as a function of position on the sky,
rather than using the magnification matrix.

From  Eq.~\eqref{eqn:lensing_deflection} we know that the deflection field is a
sum of projected gradient terms. Approximating the light path by the
unperturbed line of sight, the deflection field can be expressed as
an angular gradient. In detail, relating the projected and angular gradients at distance
$\chi$ via $\nabla_\perp\Psi = (\grad\Psi)/\chi$, we can define a lensing
potential $\pot(\vx)$ so that
\be
\vdefl(\vx) = \grad \pot( \vx ),
\ee
where the derivative is now with respect to position on the sky.
The lensing potential can be calculated as the line-of-sight integral
%
%
\be
\pot(\vx) = -2 \int_0^{\chi_*} d\chi \left( \frac{ \chi_* - \chi }{\chi_* \chi} \right) \gp(\chi \vx; \eta_0 - \chi).
\label{eqn:lensing_potential}
\ee
%

To what extent is the deflection angle approximated by
Eq.~(\ref{eqn:lensing_potential}) valid? The major assumption
was the Born approximation,
which takes the integral of Eq.~\eqref{eqn:lensing_potential} along the
unperturbed path.
This will be a good approximation if the deflection angles are
sufficiently small, or if all of the lensing occurs in a single
thin plane.
 The dependence of the lensing potential on this assumption can be investigated
quantitatively under the second-order Born approximation -- taking the gravitational potential
along the perturbed path given by the first-order result.
The effect on the gradient component of $\vdefl(\vx)$
is computed in \cite{Cooray:2002mj,Shapiro:2006em}, and is expected
to be negligible for any future experiment.
A potentially more important aspect of the second-order Born approximation is
that it introduces a non-zero curl component into the deflection field,
however this is not expected to be significant for experimental noise
levels achievable in planned future surveys (i.e.\ greater than
0.25 $\mu$K-arcmin)~\cite{Hirata:2003ka}. The validity of
the Born approximation can also potentially be investigated by ray-tracing
through numerical simulations \cite{Das:2007eu}, although this has not yet
been done.

\subsection{\textit{Lensing potential power spectrum}}

The lensing potential depends on the geometry of the universe and the gravitational potential along the line of sight,
and hence, like galaxy lensing, is a probe of cosmology. Assuming linear evolution from the primordial density fluctuations,
we can take $\gp$ in Fourier space to be given by
\be
\gp(\vk, \eta) = T_{\gp}(k; \eta) {\cal R}(\vk)\, ,
\ee
where $T_{\gp}(k; \eta)$ is the transfer function, and ${\cal R}(\vk)$
are the primordial curvature fluctuations, with 3D power spectrum ${\cal P}_{\cal R}(\vk)$.
Under this assumption, expanding $\pot(\vx)$ in harmonic space and
taking the angular power spectrum gives \cite{Lewis:2006fu}
\be
C_l^{\pot\pot} = 16 \pi \int \frac{dk}{k}
{\cal P}_{\cal R}(k) \left[ \int_0^{\chi_*} d\chi T_{\gp}(k; \eta_0 - \chi) j_l(k\chi)
\left( \frac{\chi_* - \chi}{\chi_* \chi} \right) \right]^2 \, ,
\label{eqn:alpha_ps}
\ee
where $j_l(k\chi)$ is a spherical Bessel function.
The \emph{deflection} power spectrum, $l(l+1)C_l^{\pot}$,
for a typical model is plotted in Fig.~\ref{fig:clphiphi}.
It is customary to plot $[l(l+1)]^2 C_l^{\pot\pot} /(2\pi)$ because
\be
\int d\ln l\, \frac{[l(l+1)]^2C_l^{\pot\pot}}{2\pi} \approx
\langle \vdefl \rangle^2 ,
\ee
the variance of the deflections.

The assumption of linear theory and Gaussian primordial fluctuations
implies that $\gp$ is Gaussian as well. However, on small scales it is clear
that the potentials are subject to a large degree of non-linear evolution.
Fortunately, the size threshold for non-linearity is small enough that there
are many such non-linear clumps on the way to the last-scattering surface. This Gaussianizes the projection of
$\gp$ to some extent, but the non-linearity does still appear as an enhanced
power spectrum, $C_l^{\pot\pot}$, on small scales; see Fig.~\ref{fig:clphiphi}.
Non-linearity may therefore
be approximately accounted
for by scaling the linear transfer functions according to
\be
T_{\Psi}(k, \eta) \rightarrow T_{\Psi}(k, \eta) \sqrt{ \frac{ {\cal P}_{\Psi}^{\text{non-linear}}(k; \eta)}{ {\cal P}_{\Psi}(k; \eta) } }.
\label{eqn:halofit_corrections}
\ee
The non-linear power spectrum can be calculated from analytical fitting formulae
such as those of \halofit \cite{Smith:2002dz}; this approach is implemented in
the \camb code \cite{Lewis:1999bs}.
The ultimate test of the Gaussian approximation is to propagate CMB photons through
large numerical $N$-body simulations.
Several groups have developed methods for creating the concentric-shell
mass maps necessary to perform the radial integral of
Eq.~\eqref{eqn:lensing_deflection} numerically~\cite{Das:2007eu,Carbone:2008fa,Fosalba:2007mf},
and all find that the \halofit approach works well at the percent level
for $l<2000$.

\section{Lensing effects on CMB observables}
\label{sec:lensing_effects_on_cmb_observables}

In Fig.~\ref{fig:lensed_unlensed_map} we show the difference
between the lensed and unlensed sky for a realization of the
CMB temperature. In this picture, the effect of lensing is to
create a non-Gaussian ``boiling soup'' pattern in the observed CMB. Large-scale
lenses create coherent displacements across large patches of
the sky, which appear as a modulation of the gradient of the unlensed
CMB in the difference map.

To study the properties of the lensed CMB analytically, it is often
useful to use a Taylor expansion of the lensing displacements.
With the CMB temperature, for example, we have
\ba
\tl(\vx)
&=& \tu(\vx + \vdefl(\vx)) \nn \\
&=& \tu(\vx) + \vdefl^{a}  \grad_{a} \tu(\vx) + \frac{1}{2} \vdefl^{a} \vdefl^{b} \grad_{a} \grad_{b} \tu(\vx) + \cdots
\label{eqn:lensing_expansion_t}
\ea
The first line is not strictly correct outside the flat-sky approximation
due to the curvature of the sky, however the Taylor expansion of the second
line is, provided that the gradients are treated as covariant derivatives.
This is also true for polarization provided
that the polarization tensor is used in place of $T$~\cite{Challinor:2002cd}.

The expansion in Eq.~(\ref{eqn:lensing_expansion_t})
is frequently useful to gain intuition
into lensing effects. Consider, for example, the first-order
expansion written in harmonic space:
\be
[\vdefl^a\grad_{a} \tu](\vl) = [\grad^{a} \pot \grad_{a} \tu](\vec{l}) = -\int \frac{d^2 \vec{L}}{2\pi} \vec{L}' \cdot (\vec{l} - \vec{L}) \pot(\vec{L}) \tu(\vec{l} - \vec{L}) \, .
\label{eqn:lensing_expansion_first}
\ee
Here we can see that a mode of $\pot(\vec{L})$ with wavevector $\vec{L}$
couples the unlensed CMB at wavevector $\vec{l}-\vec{L}$ into the
observed CMB at wavevector $\vec{l}$. For fixed lenses, this mode-coupling
introduces off-diagonal components into the covariance matrix of the observed
CMB, with a characteristic
spacing of $\delta{l}=50$ given by the peak of the
deflection angle
power spectrum. Specifically,
\be
\langle \tilde{T}(\vec{l}_1) \tilde{T}(\vec{l}_2) \rangle_{\text{CMB}}
= \frac{1}{2\pi}\pot(\vec{l}_1 + \vec{l}_2) \cdot \left[
\vec{l}_1 C_{l_1}^{TT} + \vec{l}_2 C_{l_2}^{TT} \right]
\qquad (\vec{l}_1 \neq \vec{l}_2)
\, ,
\ee
where the average is over an ensemble of CMB fluctuations for fixed lenses.
This is an interesting feature; large-wavelength modes of
the lensing potential introduce correlations between CMB modes on much
smaller scales.
In Sec.~\ref{sec:lens_reconstruction} we will see how
a reconstruction of the lensing potential can be made by correlating
a large number of small-scale modes.

We note that the convergence of the lensing Taylor expansion is
somewhat suspect -- the typical RMS deflection angle
of $\sim 3\arcmin$ is comparable to the $10\arcmin$ scale length of modes
at $l = 1000$, and so the expansion converges slowly
at the map level.
Nevertheless, when calculating ensemble-averaged quantities
the lensing Taylor expansion is often quite accurate. In this
case, it is relative displacements rather than the actual lensing
effects which are important, and the Taylor expansion captures
these reasonably well~\cite{Hanson:2009kg}.
However the expansion is not accurate enough for a percent-level calculation of the lensed power spectrum.

\subsection{\textit{Power spectra}}

The CMB  power spectra completely characterize the primary fluctuations for Gaussian,
statistically-isotropic models. Even though the lensed sky is slightly non-Gaussian, the lensed power spectra
are still useful for comparison with observations.
Here we discuss the calculation of the lensed power
spectra  using the Taylor-series expansion described above.
A more accurate non-perturbative result may be derived using correlation functions
\cite{Seljak:1995ve,Zaldarriaga:1998ar,Challinor:2005jy}, however
the Taylor approach gives more intuitive results and is less
mathematically involved.

\begin{figure}
\begin{center}
\includegraphics[width=3in]{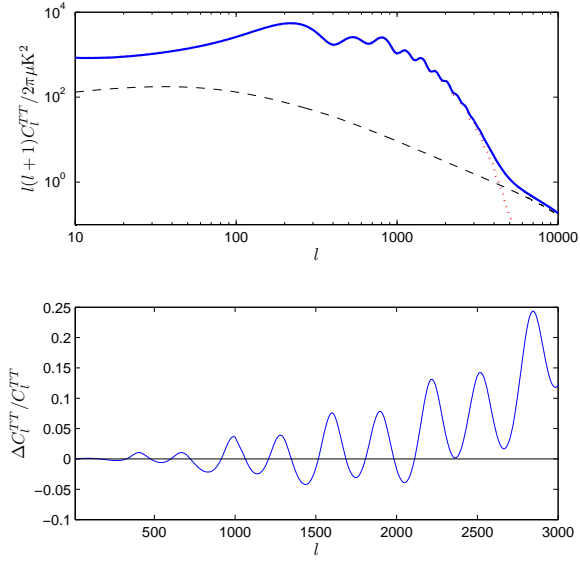}
\caption{Top: the lensed (solid) and unlensed (dotted) temperature power spectrum, as well as the small-scale approximation of Eq.~\eqref{eqn:lensedT_smallscales} (dashed). Bottom: the fractional change in the power spectrum due to lensing.
}
\label{fig:lensedT}
\end{center}
\end{figure}

We will begin with temperature. For the purposes of compact notation,
it is useful to write the lensing Taylor expansion schematically as
\be
\tl(\vl) = \tu(\vl) + \delta \tu(\vl) + \delta^2 \tu(\vl) + \cdots \, ,
\label{eqn:lensing_expansion_schematically}
\ee
where the power of delta indicates the order in the lensing potential.
The power spectrum to lowest order in $C_{l}^{\pot\pot}$ then follows from
\be
C^{\tl \tl}_{l} \delta(\vec{l}-\vec{l}') \approx
C^{\tu \tu}_{l} \delta(\vec{l}-\vec{l}') +
\langle \delta\tu(\vec{l}) \delta\tu^*(\vec{l}') \rangle + \langle \tu(\vec{l}) \delta^2 \tu^*(\vec{l'}) \rangle
+  \langle  \delta^2 \tu(\vec{l})\tu^*(\vec{l}')
\rangle \, .
\ee
In the flat-sky approximation, the gradient operations are easily taken in Fourier space: $\delta \tu(\vec{l})$ is given by Eq.~(\ref{eqn:lensing_expansion_first}) while~\cite{Hu:2000ee}
\be
\delta^2 \tu(\vec{l}) = -\frac{1}{2} \int \frac{d^2 \vec{l}_1}{2\pi} \int \frac{d^2 \vec{l}_2}{2\pi} \vec{l}_1 \cdot (\vec{l}_1 + \vec{l}_2 - \vec{l}) \vec{l}_1 \cdot \vec{l}_2 \tu(\vec{l}_1) \pot(\vec{l_2})\pot^{*}(\vec{l}_1 + \vec{l}_2 - \vec{l})\, ,
\ee
from which the power spectrum may be derived.
Assuming that both $\tu(\vec{l})$ and $\pot(\vec{l})$
have the statistics of an isotropic, Gaussian random field we find:
\ba
\langle \tu(\vec{l}) \delta^2 \tu^*(\vec{l}') \rangle
+ \langle \delta^2 \tu(\vec{l}) \tu^*(\vec{l}')
\rangle &=& -l^2 R^{\pot} C_l^{\tu \tu} \delta(\vec{l}-\vec{l}') \\
\langle \delta\tu(\vl) \delta\tu^*(\vl') \rangle &=&
\delta(\vl-\vl')
\int \frac{d^2 \vec{L}}{(2\pi)^2} \left[ \vec{L} \cdot (\vec{l} - \vec{L}) \right]^2 C^{\pot\pot}_{|\vec{l} - \vec{L} |} C^{\tu \tu}_{L}
\label{eqn:lensed_cltt} \, ,
\ea
where $R^{\pot}$ is half of the total mean-squared deflection,
\be
R^{\pot} \equiv \frac{1}{2}\langle |\grad \pot|^2 \rangle  = \frac{1}{4\pi} \int \frac{dl}{l} l^4 C^{\pot \pot}_{l} \sim 3\times10^{-7} \, .
\label{eqn:rp}
\ee
Assembling these results together, we find the lensed temperature power spectrum
to first-order in $C_l^{\pot\pot}$
\be
C_l^{\tl \tl} = (1-l^2 R^\pot) C_l^{\tu\tu} + \int \frac{d^2 \vec{l}'}{(2\pi)^2}
\left[\vec{l'}\cdot (\vec{l}-\vec{l}')\right]^2 C^{\pot\pot}_{|\vec{l}-\vec{l}'|} C_{l'}^{\tu\tu} \, .
\label{eqn:lensedCT}
\ee
The second term, which represents a convolution
of the unlensed temperature power spectrum with the lensing potential power
spectrum,
almost cancels the $R^\pot$ term over the acoustic part of the
spectrum, leaving only a small blurring of the acoustic peaks with a
characteristic kernel width of $\delta l \sim 50$.
This is illustrated in Fig.~\ref{fig:lensedT}, where we plot the lensed and unlensed spectra.

The unlensed CMB has very little power on small scales, $l>3000$, due
to diffusion damping.
Deep enough into the damping tail, any non-negligible
contribution to Eq.~\eqref{eqn:lensedCT} from the unlensed temperature power spectrum will be from modes with $l' \ll l$, and so the lensed spectrum
may simplified as~\cite{Hu:2000ee,Zaldarriaga:2000ud}
%
\be
C_l^{\tl\tl} \approx
C_l^{\pot\pot} \int \frac{d^2 \vec{l}'}{(2\pi)^2}(\vec{l'} \cdot \vec{l})^2 C_{l'}^{\tu \tu}
\approx l^2 C_l^{\pot\pot} \int \frac{dl'}{l'} \frac{(l')^4 C_{l'}^{\tu \tu}}{4\pi}
\approx l^2 C_l^{\pot\pot} R^{\tu} \, ,
\label{eqn:lensedT_smallscales}
\ee

where $R^{\tu}$ is defined analogously to $R^{\pot}$ in Eq.~\eqref{eqn:rp},
and is half the mean-squared gradient of the unlensed CMB.
 This causes
an increase in the lensed CMB power of $\sim 10\%$ at $l>2000$, as we can see in Fig.~\ref{fig:lensedT}.
Qualitatively, there are therefore two lensing effects on the temperature power spectrum:
large-scale lenses smooth the peaks of the observed power spectrum,
while small-scale lenses introduce small-scale power in the damping tail by lensing of the large-scale temperature gradient.

We now turn to the lensed polarization power spectra.
For simplicity, we will assume that $\bu(\vl) = 0$, i.e. that there are no primordial $B$-modes.
In the flat-sky limit, lensing remaps the Stokes parameters defined on a fixed
basis in the same way as for the temperature.
The calculation is therefore similar to the temperature case, with the lensed spectra to lowest
order in $C_l^{\pot\pot}$ given by \cite{Hu:2000ee}
\begin{eqnarray}
C_l^{\tl\el} &=& (1-l^2R^\pot)C_l^{\tu\eu} +
+ \int \frac{d^2 \vl'}{(2\pi)^2}\left[\vl'\cdot (\vl-\vl')\right]^2 C_{|\vl-\vl'|}^{\pot\pot} C^{\tu\eu}_{l'} \cos [2(\sr_{\vl'}-\sr_\vl)]
\label{eqn:lensed_te}
\\
C_l^{\el\el}  &=& (1-l^2R^\pot)C_l^{\eu\eu} + \int \frac{d^2 \vl'}{(2\pi)^2}\left[\vl'\cdot (\vl-\vl')\right]^2 C_{|\vl-\vl'|}^{\pot\pot} C^{\eu\eu}_{l'} \cos^2 [2(\sr_{\vl'}-\sr_\vl)]
\label{eqn:lensed_e}
\\
C_l^{\bl\bl}  &=&  \int \frac{d^2 \vl'}{(2\pi)^2}\left[\vl'\cdot (\vl-\vl')\right]^2 C_{|\vl-\vl'|}^{\pot\pot} C^{\eu\eu}_{l'}
\sin^2 [2(\sr_{\vl'}-\sr_\vl)],
\label{eqn:lensed_b}
\end{eqnarray}
where $R^{\pot}$ was defined in Eq.~\eqref{eqn:rp}. Lensing of purely $E$-modes therefore generates $B$-mode polarization, even when there is none initially.

\begin{figure}
\begin{center}
\includegraphics[width=3in]{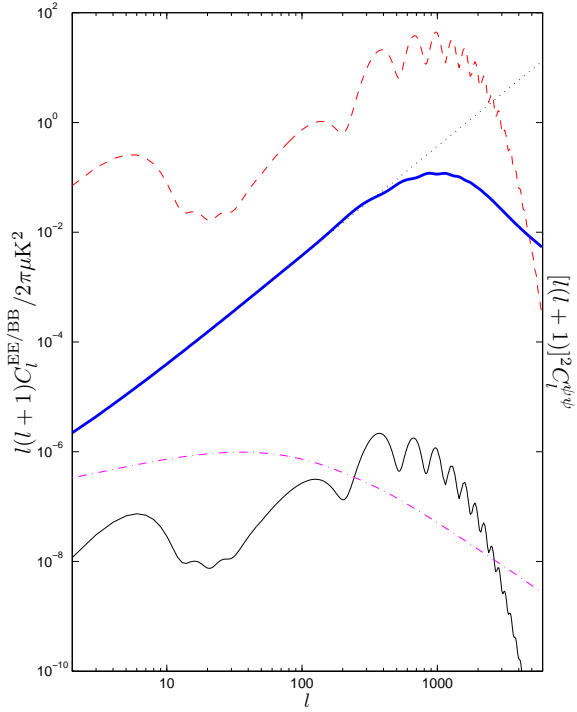}
\caption{The unlensed $E$-mode power spectrum $l(l+1)C_l^{EE}/2\pi$ (top; dashed), lensing deflection power spectrum $[l(l+1)]^2C_l^{\pot\pot}$ (dot-dashed), and the contribution to the large-scale lensed $B$-mode power spectrum from each $\log l$ given by half their product (thin solid). The dotted line is the large-scale white spectrum $l(l+1)C_l^{\bl\bl}/2\pi$ given by Eq.~\eqref{eqn:lowl_lensedb}, which can be compared to the full numerical result (thick solid). The lensed $B$-mode power spectrum has a close-to-white spectrum at $l\ll 1000$.
\label{fig:EtoB}}
\end{center}
\end{figure}

The lensing of the $E$-mode power spectrum, and the $TE$ spectrum,
is qualitatively similar
to that of temperature. There is a convolution which blurs the features
of the spectrum and shifts power into the damping tail.
Quantitatively, the magnitude of the lensing effects is
larger for the $E$-modes than for temperature, because the peaks in the
unlensed power spectrum are rather sharper, leading to larger fractional changes
in the lensed power of $\clo(30\%)$ near the acoustic peaks \cite{Zaldarriaga:1998ar,Challinor:2005jy,Hu:2000ee}.

Lensing of the CMB polarization causes power to leak from $E$- to $B$-modes,
and this may be an important contaminant in the search for primordial $B$-modes
on large angular scales.
Since the unlensed $E$-mode spectrum peaks around $l \sim 1000$ (see
Fig.~\ref{fig:EtoB}), for $l \ll 1000$ we can take
$|\vec{l}'| \gg |\vec{l}|$ in Eq.~(\ref{eqn:lensed_b}) giving
\ba
C^{\bl\bl}_{l} &\approx& \int \frac{d^2 \vl'}{(2\pi)^2} |\vl'|^4 C^{\pot\pot}_{l'} C^{\eu\eu}_{l'} \sin^2 2(\sr_{\vl'} - \sr_{\vl}) \nn \\
&=& \frac{1}{4\pi} \int \frac{d l'}{l'} {l'}^4 C_{l'}^{\pot\pot} {l'}^2 C_{l'}^{\eu\eu}.
\label{eqn:lowl_lensedb}
\ea
This is independent of $l$, corresponding to a white-noise spectrum of $B$-modes.
In Fig.~\ref{fig:EtoB}
we can see that the lensed $B$-modes are nearly white for $l<100$,
with the level predicted by Eq.~(\ref{eqn:lowl_lensedb}).
For the fiducial model we have $C_l^{\bl\bl}\sim2\times10^{-6}\,\mu \text{K}^2$,
corresponding to an instrumental white-noise level of $\sim 5\,\mu\text{K}$-arcmin. The next generation of polarization surveys should achieve noise levels
close to this and lensing will therefore become an important source of confusion
for them.

To assess the amount of additional cosmological information in the lensed
spectra over the unlensed, consider attempting to constrain departures
of $C_l^{\pot\pot}$ from some fiducial spectrum:
\be
C_l^{\pot\pot} = (1+f_l) C_l^{\pot\pot} |_{\rm fid.} \, .
\ee
An eigenmode analysis of the Fisher matrix for this problem can be used
to identify modes of $f_l$ which are constrained well by data.
For a cosmic-variance limited experiment to $l_{\rm max}=2000$,
there are essentially two such modes \cite{Smith:2006nk}.
From the $TT$ and $EE$ power spectra, there is a mode
which peaks at $l=100$ and tracks the dominant lensing power.
These are the modes of $C_l^{\pot\pot}$ which smooth the observed power spectra.
Including $B$-mode polarization, there is a second mode which is
much broader, reflecting the fact that the large scale $B$-mode
power is sourced by lenses on all scales. This mode peaks at $l=200$,
but has a broad tail extending past $l=1000$.

Increasing the $l_{\rm max}$ of the temperature measurement beyond $l=2000$,
the small-scale modes
of the lensing potential would ideally also be constrained
by the increase in the small-scale lensed power which they cause. However,
in this regime extra-Galactic foregrounds and secondary effects will
provide a source of confusion which is, as yet, not well understood.

The amplitude of the well-constrained $C_l^{\pot\pot}$ modes is somewhat dependent on cosmology,
and so they may be used to place improved constraints on parameters.
Some care is required with the $B$-mode power spectrum due to the
highly non-Gaussian nature of the lens-induced $B$-modes leading to
correlations between estimates of $C_l^{\bl\bl}$ at different scales~\cite{Smith:2006nk,Smith:2005ue,Li:2006pu}. On the whole, the power spectra are a
rather sub-optimal statistic for extracting information from CMB lensing, however. Improved constraints
can be made with estimators which are designed specifically to detect the lensing signal;
these are discussed in Sec.~\ref{sec:lens_reconstruction}.

\subsection{\textit{Bispectra}}
\label{sec:lensing_effects_on_cmb_observables_bispectra}

The bispectrum, or three-point correlation function is given by
\be
\langle \tl(\vl_1) \tl(\vl_2) \tl(\vl_3) \rangle = \tilde{B}(\vl_1, \vl_2, \vl_3)
\delta(\vl_1+\vl_2+\vl_3) \, .
\ee
It is the simplest, lowest-order statistic which can be used to test for
a skewed non-Gaussianity of the CMB fluctuations
and so it has received
much attention (see~\cite{Bartolo:2004if} for a review).
Considering the effects to second order in the
lensing potential, the lensed bispectrum is given in terms
of the unlensed bispectrum by
\ba
\tilde{B}(\vl_1, \vl_2, \vl_3) \delta(\vl_1+\vl_2+\vl_3) &=& B(\vl_1, \vl_2, \vl_3)
\delta(\vl_1+\vl_2+\vl_3)
 + \Big( \langle \tu(\vl_1) \tu(\vl_2) [\grad \pot \cdot \grad \tu] (\vl_3) \rangle \nn \\
&+& \langle \tu(\vl_1) [\grad \pot \cdot \grad \tu] (\vl_2)
[\grad \pot \cdot \grad \tu] (\vl_3) \rangle \nn \\
&+& \frac{1}{2} \langle \tu(\vl_1) \tu(\vl_2) [\nabla^a \pot \nabla^b \pot
\nabla_a T \nabla_b T ](\vl_3)
\rangle + \mbox{2 perms}
\Big) \, .\label{eqn:lensed_bispectrum}
\ea
One effect of the lensing terms is to
generate a smoothing of
features in the primordial
bispectrum analogous to the effect
in the power spectra \cite{Cooray:2008xz,Hanson:2009kg}.
This effect is not significant enough to affect the analysis
of CMB data in the near future however \cite{Hanson:2009kg}.

A more important effect is due to the correlation of the CMB
lensing potential with the large-scale temperature anisotropies.
%
The CMB radiation we observe is redshifted or blueshifted by the evolution
of potentials along the line of sight, a phenomenon often referred to as the (late-time) integrated Sachs-Wolfe (ISW) effect
\cite{1967ApJ...147...73S,1968Natur.217..511R}.
The ISW contribution to the CMB anisotropies from linear potentials is significant at low multipoles ($\ell < 100$)
due to the global effects of dark energy,
but dies off rapidly on smaller scales.
Since the evolving potentials that generate the ISW effect also cause lensing, there is some cross-correlation $C^{\tu \pot}$ between
$\tu$ and $\pot$, which is calculated in e.g. \cite{Hu:2000ee}.
The correlation can be neglected for calculating the lensed power spectra, however
a non-zero $C_l^{\tu \pot}$ generates a bispectrum, by the second term in Eq.~(\ref{eqn:lensed_bispectrum}), even if the primordial fluctuations are Gaussian~\cite{Zaldarriaga:2000ud,Goldberg:1999xm}.
At leading order in $\pot$ this is given by
\be
\langle \tl(\vec{l}_1) \tl(\vec{l}_2) \tl(\vec{l}_3) \rangle
\approx - \frac{1}{2\pi} \delta(\vec{l}_1 + \vec{l}_2 + \vec{l}_3)
(C^{\tu \pot}_{l_1} C^{\tu\tu}_{l_2} \vec{l}_1 \cdot \vec{l}_2 + \mbox{5 perms.})\,
.
\ee
The value of this induced
non-Gaussianity is greatest for ``squeezed'' triangles in which
$l_1 \sim l_2 \gg l_3$ (or permutations thereof), corresponding to a large-scale temperature mode which is sourced by the ISW effect and two small-scale
modes which are affected by lensing. This closely
mimics the local form of non-Gaussianity which is predicted
in a range of possible early-Universe scenarios (for a review, see~\cite{Fergusson:2008ra}).
For upcoming experiments such as Planck, this lensing-induced
bispectrum will provide a large source of contamination for
the measurement of primordial non-Gaussianity, and will need
to be accurately accounted for to extract the primordial bispectrum~\cite{Hanson:2009kg,Serra:2008wc}.

Note that because the lensing effects are linear in the observed
CMB, lensing cannot generate a non-zero bispectrum in the absence
of the $\tu \pot$ correlation, and the lowest order at which the
lensing induces non-Gaussianity is the four-point correlation
function or trispectrum. This has been studied thoroughly by several authors
\cite{Hu:2001fa,Kesden:2003cc}.

\section{Lens reconstruction}
\label{sec:lens_reconstruction}

In the previous section we discussed
how lensing affects the CMB power spectra and temperature bispectrum.
We turn now to the issue of extracting more information about the lensing itself,
in particular how to construct an optimal estimator for the lensing
deflection field from the observed CMB. This reconstruction can
then be connected to cosmological parameters with a likelihood analysis.

If we assume that the lensing potential is fixed then
the statistics of the lensed CMB remain
Gaussian, however the inhomogeneities which the lenses introduce
make it statistically anisotropic. In Fourier space, this manifests itself as
correlations between modes with $\vl \ne \vl'$, as we have already
discussed. In real space, the manifestation is somewhat more intuitive: for an isotropic model,
the correlation between two points depends only on the distance between them, and
the effect of the lens remapping is to make the observed, lensed distance between
two points slightly larger or smaller than the primary, unlensed distance on which
the correlation is defined.
This feature may be exploited to produce estimators for the lensing potential.
Consider, for example, the lensing expansion in Eq.~\eqref{eqn:lensing_expansion_t},
which shows that at first order in $\pot$ the effect of lensing is
to introduce a dependence into the observed CMB
on the gradient of the unlensed CMB.
This suggests that a quadratic estimator for the deflection angle of the form
\be
\widehat{\vdefl}(\vx) \propto \tl(\vx) \grad{\tu}(\vx)
\label{eqn:heuristic_estimator}
\ee
can be used to extract a measure of the lensing signal,
if properly normalized. Taking the expectation value over
realizations of the CMB, this estimator
would recover the deflection field in the mean (at least
to first order in the lensing expansion). The reconstruction
is clearly quite noisy, however, as for any particular realization
there will be some spurious $\tu \grad\tu$ correlation in the unlensed CMB.
Additionally, it is not clear how we would obtain the unlensed
CMB $\tu(\vx)$ with which to perform the correlation.
It is critical that the gradient of the unlensed CMB be used in
Eq.~(\ref{eqn:heuristic_estimator}) -- if instead the gradient of the
lensed CMB is used, the expectation value over CMB realizations vanishes
to first order in $\psi$.

Here we will derive a more rigorous lensing estimator, by
maximizing the likelihood of $\reconvar(\vx)$ given the observed data,
loosely following the approach of \cite{Hirata:2002jy}.
We will take the measurement to consist of noisy maps
of $T$, $Q$ and $U$ measurements (or some subset thereof). We combine the
observables at each point $\vx$ into a vector $\vao(\vx) = \val(\vx) + \vnse(\vx)$, where $\vnse$ is the instrumental noise realization, which is also assumed
to be Gaussian.
%
The log-likelihood $\lp$ of the observed CMB is then given by
\be
-\lp(\vao | \reconvar) = \frac{1}{2} \vao^{T} \left( \mobscov\right)^{-1} \vao + \frac{1}{2} \ln \det(\mobscov),
\ee
where $\mobscov(\vx,\vy) = \mlensedcov(\vx,\vy) + \mnoisecov(\vx,\vy)$ is the covariance of the observation (including the effects of lensing) and we have left the
integrals over positions implicit.
We will construct an estimator for the deflection field by maximizing the likelihood with respect to the deflection field
$\reconvar(\vx)$. Taking the functional derivative of
the likelihood with respect to the vector components $\reconvar_i$ we find
\be
\frac{\delta \lp}{\delta \reconvar_i(\vx)} =
\frac{1}{2} \vao^{T} (\mobscov)^{-1} \frac{\delta \mobscov}{\delta \reconvar_i(\vx)} (\mobscov)^{-1} \vao
-\frac{1}{2} \tr \left[ (\mobscov)^{-1} \frac{\delta \mobscov}{\delta \reconvar_i(\vx)} \right].
\ee
The trace term results in a ``mean field'' over realizations
of the unlensed CMB and noise. To see this, consider the identity
$\tr( \mat{A} ) = \langle \vx^{T} \mat{A} \mcov^{-1} \vx \rangle$,
where $\mat{A}$ is any matrix and $\vx$ is a vector of
Gaussian random variables with covariance $\mcov$.
Making this substitution with $\mcov = \mobscov$
and maximizing the likelihood by setting
$\delta \lp / \delta \reconvar_i(\vx) = 0$ gives the simple equation
\be
\frac{\delta \lp}{\delta \reconvar_i(\vx)} = \vhq_i(\vx)
- \langle \vhq_i(\vx) \rangle = 0 ,
\label{eqn:max_like_cond}
\ee
where the components $\vhq_i(\vx)$ are given by
\be
\vhq_i(\vx) = \frac{1}{2} \left[ (\mobscov)^{-1} \vao \right]^{T} \frac{\delta \mlensedcov}{\delta \reconvar_i(\vx)} \left[ (\mobscov)^{-1} \vao \right].
\label{eqn:vhq}
\ee
The values of $\delta \mlensedcov / \delta \reconvar_i(\vx)$ are calculated analytically
from the $\clo(\reconvar)$ covariance
in e.g. \cite{Hu:2001kj,Okamoto:2003zw}
(albeit with a slightly different notational convention than the one used here).
Equation~\eqref{eqn:max_like_cond} may be solved
iteratively using the Newton-Raphson method by updating $\reconvar(\vx)$
according to
\be
\delta\hreconvar_{i}(\vx) = - \int d\vy\,\eal_{ij}(\vx,\vy) [\vhq_j
- \langle \vhq_j \rangle](\vy) ,
\label{eqn:lensing_newton_raphson}
\ee
where sub-scripted quantities are evaluated at the current
estimate of $\reconvar(\vx)$,
and the normalization $\eal_{ij}(\vx,\vy)$ is given by
\be
\eal_{ij}(\vx,\vy) =  \left[\frac{\delta^2 \lp}{\delta \reconvar_i(\vx)\delta
\reconvar_j(\vy)}\right]^{-1} =
\left[\frac{\delta}{\delta \reconvar_i(\vx)} (\vhq_j-\langle \vhq_j \rangle)(\vy) \right]^{-1},
\ee
which in practice can be approximated by the Fisher matrix (see below).
It is apparent
that the quadratic term $\vhq(\vx)$ produces a
measure of the \reconvartext~$\reconvar(\vx)$.
Note that in the absence of lensing, with homogeneous sky coverage we must have $\langle \vhq_i(\vx) \rangle = 0$, as the expectation value would otherwise pick out a preferred direction on the sky~\cite{Hirata:2002jy}.

In fact, $\vhq_i(\vx)$ is quite
closely related to the heuristic estimator of Eq.~\eqref{eqn:heuristic_estimator}.
To show this, we begin by finding
the derivative of the lensed CMB with respect to the deflection angle.
This is straightforward since $\val(\vx) = \vau(\vx + \vdefl(\vx))$, so
\be
\frac{\delta}{\delta \vdefl_i(\vy)} \val(\vx) = \delta(\vx-\vy) (\grad_i \vau)(\vx + \vdefl(\vx)),
\ee
i.e. if the deflection angle is changed at a point, the lensed field only
changes at the same point and by an amount that depends on the gradient of the unlensed fields $\grad \vau$ at the
source
position.
To proceed further, it is useful to define the linear lensing operator $\lop(\vx,\vy)$
such that the lensed CMB is given by $\val(\vx) = \vau(\vx +\vdefl(\vx))=\int d^2 \vy \lop(\vx,\vy) \vau(\vy)$, or explicitly
\be
\lop(\vx,\vy) = \delta(\vx + \vdefl(\vx)-\vy).
\ee
With this notation the derivative of the lensed covariance $\mlensedcov = \lop \munlensedcov \lop^T$ (where $\lop^T(\vx,\vy)=\lop(\vy,\vx)$)
is given by
\be
\frac{\delta}{\delta \vdefl_i(\vx)}  \mlensedcov(\vy,\vz) = \delta(\vx-\vy) \int d^2\vy' d^2\vz'  \lop(\vy,\vy')  \grad_{y^{'i}} \munlensedcov(\vy',\vz')\lop(\vz,\vz') +  (\vy\leftrightarrow\vz).
\ee
Equation~\eqref{eqn:vhq} then simplifies to a real-space operation on filtered fields
\cite{Hirata:2002jy}:
\ba
\vhq_i(\vx)  &=& \vF_1^T(\vx) (\grad_i \vF_2)(\vx+\vdefl(\vx)),
\label{eqn:vhq2}
\ea
where the filtered fields are defined by
\be
\vF_1(\vx) \equiv  [(\mobscov)^{-1}\vao](\vx) \, ,
\qquad \vF_2(\vx) \equiv  [\munlensedcov \lop^T (\mobscov)^{-1}\vao](\vx) \, .
\ee
The filtered field $\vF_2 = \lop^{-1}\mlensedcov (\mobscov)^{-1}\vao$
is the Wiener reconstruction of the unlensed CMB, i.e.\ the optimal estimate
of the unlensed CMB given the observed noisy data, the estimate of $\reconvar$
and the unlensed CMB covariance matrix.
The gradient term $(\grad \vF_2)(\vx+\vdefl(\vx))$ is then just the best estimate of the unlensed gradient evaluated at the unlensed position, and this is contracted with
the inverse-covariance-weighted lensed CMB at $\vx$ to form $\vhq_i(\vx)$.
The weighting in $\vF_1$ is typical of any optimal estimator of CMB observables
and serves to down-weight the estimate in areas of high instrumental noise or cosmic variance.
To understand why the $\grad \vF_2$ term is evaluated at the displaced position, let
the true deflection field differ from the current estimate $\reconvar(\vx)$
by $\delta \vdefl(\vx)$.
To linear order in $\delta \vdefl(\vx)$, the lensed CMB at $\vx$ is
\be
\val(\vx) = \vau(\vx + \reconvar(\vx)) + \delta \vdefl^i(\vx) (\nabla_i \vau)(\vx
+ \reconvar(\vx)) \, .
\ee
It is therefore necessary to multiply by the gradient of the unlensed CMB at $\vx + \reconvar(\vx)$ to obtain an estimator which is proportional to the desired
\emph{change} in deflection field when averaged over realizations of the unlensed CMB.



We now consider the error on the reconstruction.
The Fisher matrix,
\be
\f_{ij}(\vx,\vy)
= - \left\langle \frac{\delta^2 \lp}{\delta \reconvar_i(\vx) \delta \reconvar_j(\vy)} \right\rangle
= \frac{1}{2} \text{Tr} \left[ (\mobscov)^{-1} \frac{\delta \mlensedcov}{\delta
\alpha_i(\vx)} (\mobscov)^{-1} \frac{\delta \mlensedcov}{\delta
\alpha_j(\vy)} \right] \, ,
\label{eqn:fisher_def}
\ee
provides a lower bound on the variance of an unbiased
estimate of $\reconvar$ by the Cramer-Rao inequality.
The expectation value in Eq.~(\ref{eqn:fisher_def}) is
taken over realizations of the CMB and noise for fixed lenses.
This may be evaluated analytically at $\reconvar=0$ if we have
isotropic sky coverage, in which case we find
\begin{multline}
\f^{(0)}_{ij}(\vx,\vy) = \\\int d^2 \vx' d^2 \vy'  \,
\text{Tr} \left[
(\mobscov)^{-1}(\vx',\vx) \nabla_{x^i} \munlensedcov(\vx,\vy') \right.
(\mobscov)^{-1}(\vy',\vy)\nabla_{y^j}\munlensedcov(\vy,\vx')
\\+  (\mobscov)^{-1}(\vy,\vx) \nabla_{x^i} \munlensedcov(\vx,\vy')
\left.
 (\mobscov)^{-1}(\vy',\vx')\nabla_{y^j}\munlensedcov(\vx',\vy) \right]_{\alpha=\vec{0}}
\, .
\label{eqn:lens_rec_noise}
\end{multline}
%
%
This is diagonal in Fourier space in its spatial arguments; for example, for the
temperature alone we have
\ba
\f^{(0)}_{ij}(\vx,\vy) &=& \int \frac{d^2 \vL}{(2\pi)^2} \, e^{i \vL\cdot(\vx-\vy)}
N_{ij}^{-1}(\vL) \nn \\
\text{where}\,\, N_{ij}^{-1}(\vL) &\equiv& \frac{1}{2}\int \frac{d^2 \vl}{(2\pi)^2} \,  \frac{[(\vL-\vl)C^{TT}_{|\vl-\vL|}+
\vl C^{TT}_l]_i}{(C^{TT}_l + N_l)}
\frac{[(\vL-\vl)C^{TT}_{|\vl-\vL|}+ \vl C^{TT}_l]_j}{(C^{TT}_{|\vl-\vL|} + N_{|\vl-\vL|})} \, .
\label{eqn:fisher_alpha}
\ea
Here, $N_l$ is the noise power spectrum of the temperature map. To make contact
with previous work~\cite{Hirata:2003ka,Hu:2001kj} that reconstructed the lensing
potential $\psi$ rather than the deflection field, we note that the Fisher matrix
$\f_{\psi}^{(0)}$ is related to that for $\alpha$ by
$\f_{\psi}^{(0)}(\vx,\vy) = \nabla_{y^j} \nabla_{x^i} \f^{(0)}_{ij}(\vx,\vy)$.
Equation~(\ref{eqn:fisher_alpha}) then gives
\be
\f^{(0)}_{\psi}(\vx,\vy) = \int \frac{d^2 \vL}{(2\pi)^2} \, e^{i \vL\cdot(\vx-\vy)}
\left[\frac{1}{2} \int \frac{d^2 \vl}{(2\pi)^2} \,  \frac{[\vL\cdot(\vL-\vl)C^{TT}_{|\vl-\vL|}+
\vL\cdot\vl \, C^{TT}_l]^2}
{(C^{TT}_l + N_l)(C^{TT}_{|\vl-\vL|} + N_{|\vl-\vL|})} \right] \, ,
\label{eqn:psi_noise}
\ee
which agrees with the power spectrum of the reconstruction noise derived for the
optimal quadratic estimator of $\psi$
in~\cite{Hu:2001kj} in the limit of no lensing.
Note that in reconstructing $\reconvar$ rather than $\psi$ we allow for
the possibility of a curl mode in $\reconvar$. This can act as a monitor
for systematic effects, as it is expected to be zero cosmologically.
However constraints using the iterated estimator can be improved by using a prior that it is zero~\cite{Hirata:2003ka}.

%

\begin{figure}
\begin{center}
\psfrag{el}{$\ell$}
\psfrag{BIGLLPONEMULCLPP}{$[\ell(\ell+1)]^2 C_{\ell}^{\psi\psi} / 2\pi$}
\psfrag{Planck}{Planck}
\psfrag{CMBPol}{CMBPol}
\includegraphics[width=4.5in]{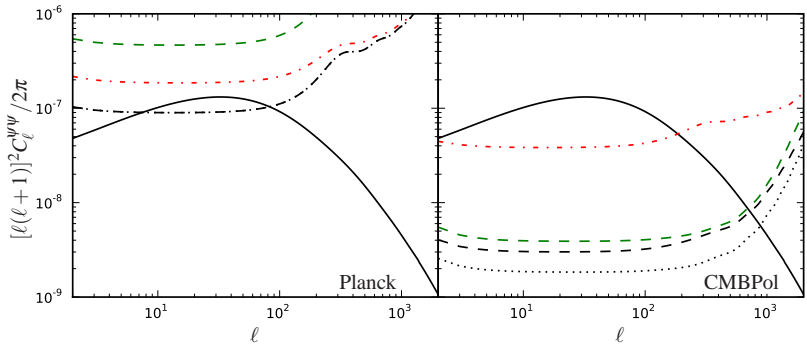}
\caption{
Power spectrum of the errors in the gradient part
of the reconstructed deflection field for two full-sky experiments.
Left: an approximation to the Planck satellite, with
$\sigma(T)=27\,\mu \mbox{K-arcmin}$.
Right: a version of the proposed CMBPol mission, with
$\sigma(T)=1\,\mu \mbox{K-arcmin}$.
For both experiments, we assume a Gaussian beam of FWHM $7\arcmin$
and $Q$ and $U$ noise levels equal to $\sqrt{2}\sigma(T)$.
The curves are the theory deflection power (solid black),
reconstruction errors from temperature alone (red dot-dashed),
polarization alone (green dashed), temperature and polarization together (black dashed),
and the Fisher limit (black dotted). For Planck, the Fisher limit is saturated.
}
\label{fig:lens_rec_noise_1_14}
\end{center}
\end{figure}

We will refer to the Fisher noise level at $\reconvar=\vec{0}$
as the ``Fisher limit''; it is plotted for two instrumental noise levels in Fig.~\ref{fig:lens_rec_noise_1_14}.
In practice, the reconstruction noise can be larger than
the Fisher limit due to the confusion created by
the lenses themselves, which act as a source of noise
in the observed CMB
and make the lensing likelihood
slightly non-Gaussian \cite{Hirata:2003ka}. For temperature
and $E$-mode polarization reconstruction these effects are generally negligible,
as the lensing represents a small perturbation to the primary signal.
For a reconstruction including $B$-mode polarization measurements, however,
the difference is very important at low noise levels as the lensing contributions
constitute a significant fraction of the observed $B$-mode power.
The true reconstruction noise can be determined from simulations \cite{Hirata:2003ka}.
In \cite{Smith:2008an} it is shown that a good approximation to
the true error may be obtained by treating the contribution
of poorly constrained modes of $\psi$ to be Gaussian. This is
done by replacing the unlensed power spectra in the denominator of
Eq.~\eqref{eqn:psi_noise} with the
spectra of the partially de-lensed CMB.
The amount of de-lensing which is possible is in turn determined by the lens
reconstruction noise (and so the reconstruction noise calculation must be iterated until convergence).
We follow this approach to calculate
the noise curves in Fig.~\ref{fig:lens_rec_noise_1_14}.

On small scales CMB lensing can also be used to study individual clusters;
here the unlensed CMB is nearly a pure gradient, and the cluster signal is a
small ``wiggle'' in this gradient due to the deflection by the cluster.
However the signal-to-noise on single clusters is generally low, and for
temperature lensing there are several sources of confusion (see, for example, Refs.~\cite{Seljak:1999zn,Yoo:2008bf}).

\section{Experimental status and prospects}

The experimental evidence for CMB lensing has developed greatly in the past several years.
At the power spectrum level, lensing causes $\clo(10\%)$ effects at $l>1000$ which are
now becoming testable with small-scale experiments.
The ACBAR team have fit for the amplitude of lensing effects
in their parameter analysis of ACBAR and WMAP data
and find a peak within one-sigma of the expected value
(although the lensing amplitude is still consistent with zero at the percent
level) \cite{Reichardt:2008ay}.
An optimal analysis of lensing effects should produce a more significant detection,
but requires stringent control of instrumental systematic effects and foregrounds.
Thus far, such an analysis has only been performed for the temperature data of the WMAP satellite.
The WMAP data only extends to $l_{\rm max} \sim 500$, and so does not contain
enough small-scale data to reconstruct the lensing potential with good
signal-to-noise. By cross-correlating a lens reconstruction from the WMAP data with a lower-noise
tracer of the large-scale structure (LSS), however, it is possible to obtain an overall detection.
This was first attempted by \cite{Hirata:2004rp}, who found no detection of lensing
effects in cross-correlation with the luminous red galaxies (LRGs) of the Sloan Digital Sky Survey (SDSS).
Their result was improved upon by \cite{Smith:2007rg}, who used radio sources from the NRAO VLA Sky Survey (NVSS),
which has a better redshift overlap with the structures which lens the CMB. They found a $3.4\sigma$ detection of lensing.
Finally, \cite{Hirata:2008cb} have improved their original analysis by incorporating correlation with NVSS sources
as well as quasars from the SDSS survey. They use a more conservative cut of the NVSS sources and a slightly
sub-optimal estimator compared to \cite{Smith:2007rg}, obtaining evidence for the expected lensing-LSS correlation
of $2.5\sigma$.
All three of these cross-correlation results have incorporated very thorough tests for both
instrumental and astrophysical systematics, most of which have negligible effects on the current overall
error budget.

At present, therefore, there is experimental evidence at modest
significance for the effect of gravitational lensing in the CMB.
High-resolution experiments which are currently taking data, such as
Planck,
ACT, and
SPT, are expected to improve
this situation dramatically, with reconstruction signal-to-noise approaching unity for many modes of the $\pot$ field. This will
enable CMB lensing to become a useful tool for precision cosmology. With lens reconstruction
from the Planck data, for example, constraints on the energy density of massive neutrinos
are expected to improve by a factor of two
over an analysis which neglects a lensing reconstruction
\cite{Kaplinghat:2003bh,Lesgourgues:2005yv,Perotto:2006rj}.
CMB lensing measurements will also
enable definitive testing between some theoretical models of the
anomalous CMB ``cold spot'' \cite{Das:2008es}.
Upcoming large-scale polarization experiments, such as BICEP2, QUIET, and PolarBeaR, are also expected
to detect lensing in polarization through the $B$-modes which it induces.

As CMB lensing moves from a purely theoretical study to an observational one, there
are a number of challenges to be overcome. The effects of foregrounds, such as
the Sunyaev-Zel'dovich effect and residual extra-Galactic point-sources, will become increasingly important
as high signal-to-noise detections of lensing are attempted. The construction of a realistic
lensing likelihood for the purpose of parameter constraints has also yet to
be demonstrated.
In the more distant future, CMB lensing should provide the dominant source of confusion
for primordial $B$-mode measurements targeting the signal from recombination
around $l \sim 100$ (for sufficiently low tensor amplitude).
Lens reconstruction with polarization and delensing will become important tools to mitigate this confusion,
and preliminary work on these techniques and their sensitivity to foregrounds and instrumental
systematics is currently under active development \cite{Smith:2008an}.


\begin{acknowledgements}
DH is supported by a Gates' Scholarship. AL acknowledges an STFC
Advanced Fellowship.
\end{acknowledgements}

\newcommand{\doi}[1]{}
\bibliographystyle{spphys}       
\bibliography{grg_review}         

\begin{thebibliography}{10}
\providecommand{\url}[1]{{#1}}
\providecommand{\urlprefix}{URL }
\expandafter\ifx\csname urlstyle\endcsname\relax
  \providecommand{\doi}[1]{DOI \discretionary{}{}{}#1}\else
  \providecommand{\doi}{DOI \discretionary{}{}{}\begingroup
  \urlstyle{rm}\Url}\fi

\bibitem{2002ApJ...581..817F}
D.J. {Fixsen}, J.C. {Mather}, Astrophys. J. \textbf{581}, 817 (2002).
\newblock \doi{10.1086/344402}

\bibitem{Hu:1997hv}
W.~Hu, M.J. White, New Astron. \textbf{2}, 323 (1997).
\newblock \eprint{astro-ph/9706147}.
\newblock \doi{10.1016/S1384-1076(97)00022-5}

\bibitem{Knox:2002pe}
L.~Knox, Y.S. Song, Phys. Rev. Lett. \textbf{89}, 011303 (2002).
\newblock \eprint{astro-ph/0202286}.
\newblock \doi{10.1103/PhysRevLett.89.011303}

\bibitem{Seljak:2003pn}
U.~Seljak, C.M. Hirata, Phys. Rev. \textbf{D69}, 043005 (2004).
\newblock \eprint{astro-ph/0310163}.
\newblock \doi{10.1103/PhysRevD.69.043005}

\bibitem{Kaplinghat:2003bh}
M.~Kaplinghat, L.~Knox, Y.S. Song, Phys. Rev. Lett. \textbf{91}, 241301 (2003).
\newblock \eprint{astro-ph/0303344}.
\newblock \doi{10.1103/PhysRevLett.91.241301}

\bibitem{Lesgourgues:2005yv}
J.~Lesgourgues, L.~Perotto, S.~Pastor, M.~Piat, Phys. Rev. \textbf{D73}, 045021
  (2006).
\newblock \eprint{astro-ph/0511735}.
\newblock \doi{10.1103/PhysRevD.73.045021}

\bibitem{dePutter:2009kn}
R.~de~Putter, O.~Zahn, E.V. Linder, Phys. Rev. \textbf{D79}, 065033 (2009).
\newblock \eprint{0901.0916}.
\newblock \doi{10.1103/PhysRevD.79.065033}

\bibitem{Partridge3K1995}
B.~Partridge, \emph{3K: The Cosmic Microwave Background Radiation}.
\newblock Cambridge Astrophysics Series (Cambridge University Press, 1995)

\bibitem{Seljak:1996gy}
U.~Seljak, M.~Zaldarriaga, Phys. Rev. Lett. \textbf{78}, 2054 (1997).
\newblock \eprint{astro-ph/9609169}.
\newblock \doi{10.1103/PhysRevLett.78.2054}

\bibitem{Kamionkowski:1996zd}
M.~Kamionkowski, A.~Kosowsky, A.~Stebbins, Phys. Rev. Lett. \textbf{78}, 2058
  (1997).
\newblock \eprint{astro-ph/9609132}.
\newblock \doi{10.1103/PhysRevLett.78.2058}

\bibitem{Lewis:2006fu}
A.~Lewis, A.~Challinor, Phys. Rept. \textbf{429}, 1 (2006).
\newblock \eprint{astro-ph/0601594}.
\newblock \doi{10.1016/j.physrep.2006.03.002}

\bibitem{Cooray:2002mj}
A.~Cooray, W.~Hu, Astrophys. J. \textbf{574}, 19 (2002).
\newblock \eprint{astro-ph/0202411}.
\newblock \doi{10.1086/340892}

\bibitem{Shapiro:2006em}
C.~Shapiro, A.~Cooray, JCAP \textbf{0603}, 007 (2006).
\newblock \eprint{astro-ph/0601226}

\bibitem{Hirata:2003ka}
C.M. Hirata, U.~Seljak, Phys. Rev. \textbf{D68}, 083002 (2003).
\newblock \eprint{astro-ph/0306354}.
\newblock \doi{10.1103/PhysRevD.68.083002}

\bibitem{Das:2007eu}
S.~Das, P.~Bode, Astrophys. J. \textbf{682}, 1 (2008).
\newblock \eprint{0711.3793}

\bibitem{Smith:2002dz}
R.E. Smith, et~al., Mon. Not. Roy. Astron. Soc. \textbf{341}, 1311 (2003).
\newblock \eprint{astro-ph/0207664}.
\newblock \doi{10.1046/j.1365-8711.2003.06503.x}

\bibitem{Lewis:1999bs}
A.~Lewis, A.~Challinor, A.~Lasenby, Astrophys. J. \textbf{538}, 473 (2000).
\newblock \eprint{astro-ph/9911177}.
\newblock \doi{10.1086/309179}

\bibitem{Carbone:2008fa}
C.~Carbone, C.~Baccigalupi, M.~Bartelmann, S.~Matarrese, V.~Springel, Mon. Not.
  Roy. Astron. Soc. \textbf{396}, 668 (2009).
\newblock \eprint{0810.4145}.
\newblock \doi{10.1111/j.1365-2966.2009.14746.x}

\bibitem{Fosalba:2007mf}
P.~Fosalba, E.~Gaztanaga, F.~Castander, M.~Manera, Mon. Not. Roy. Astron. Soc.
  \textbf{391}, 435 (2008).
\newblock \eprint{0711.1540}

\bibitem{Challinor:2002cd}
A.~Challinor, G.~Chon, Phys. Rev. \textbf{D66}, 127301 (2002).
\newblock \eprint{astro-ph/0301064}.
\newblock \doi{10.1103/PhysRevD.66.127301}

\bibitem{Hanson:2009kg}
D.~Hanson, K.M. Smith, A.~Challinor, M.~Liguori, Phys. Rev. \textbf{D80},
  083004 (2009).
\newblock \eprint{0905.4732}.
\newblock \doi{10.1103/PhysRevD.80.083004}

\bibitem{Seljak:1995ve}
U.~Seljak, Astrophys. J. \textbf{463}, 1 (1996).
\newblock \eprint{astro-ph/9505109}.
\newblock \doi{10.1086/177218}

\bibitem{Zaldarriaga:1998ar}
M.~Zaldarriaga, U.~Seljak, Phys. Rev. \textbf{D58}, 023003 (1998).
\newblock \eprint{astro-ph/9803150}.
\newblock \doi{10.1103/PhysRevD.58.023003}

\bibitem{Challinor:2005jy}
A.~Challinor, A.~Lewis, Phys. Rev. \textbf{D71}, 103010 (2005).
\newblock \eprint{astro-ph/0502425}.
\newblock \doi{10.1103/PhysRevD.71.103010}

\bibitem{Hu:2000ee}
W.~Hu, Phys. Rev. \textbf{D62}, 043007 (2000).
\newblock \eprint{astro-ph/0001303}.
\newblock \doi{10.1103/PhysRevD.62.043007}

\bibitem{Zaldarriaga:2000ud}
M.~Zaldarriaga, Phys. Rev. \textbf{D62}, 063510 (2000).
\newblock \eprint{astro-ph/9910498}.
\newblock \doi{10.1103/PhysRevD.62.063510}

\bibitem{Smith:2006nk}
K.M. Smith, W.~Hu, M.~Kaplinghat, Phys. Rev. \textbf{D74}, 123002 (2006).
\newblock \eprint{astro-ph/0607315}.
\newblock \doi{10.1103/PhysRevD.74.123002}

\bibitem{Smith:2005ue}
S.~Smith, A.~Challinor, G.~Rocha, Phys. Rev. \textbf{D73}, 023517 (2006).
\newblock \eprint{astro-ph/0511703}.
\newblock \doi{10.1103/PhysRevD.73.023517}

\bibitem{Li:2006pu}
C.~Li, T.L. Smith, A.~Cooray, Phys. Rev. \textbf{D75}, 083501 (2007).
\newblock \eprint{astro-ph/0607494}.
\newblock \doi{10.1103/PhysRevD.75.083501}

\bibitem{Bartolo:2004if}
N.~Bartolo, E.~Komatsu, S.~Matarrese, A.~Riotto, Phys. Rept. \textbf{402}, 103
  (2004).
\newblock \eprint{astro-ph/0406398}.
\newblock \doi{10.1016/j.physrep.2004.08.022}

\bibitem{Cooray:2008xz}
A.~Cooray, D.~Sarkar, P.~Serra, Phys. Rev. \textbf{D77}, 123006 (2008).
\newblock \eprint{0803.4194}.
\newblock \doi{10.1103/PhysRevD.77.123006}

\bibitem{1967ApJ...147...73S}
R.K. {Sachs}, A.M. {Wolfe}, Ap. J \textbf{147}, 73 (1967).
\newblock \doi{10.1086/148982}

\bibitem{1968Natur.217..511R}
M.J. {Rees}, D.W. {Sciama}, Nature \textbf{217}, 511 (1968).
\newblock \doi{10.1038/217511a0}

\bibitem{Goldberg:1999xm}
D.M. Goldberg, D.N. Spergel, Phys. Rev. \textbf{D59}, 103002 (1999).
\newblock \eprint{astro-ph/9811251}.
\newblock \doi{10.1103/PhysRevD.59.103002}

\bibitem{Fergusson:2008ra}
J.R. Fergusson, E.P.S. Shellard, Phys. Rev. \textbf{D80}, 043510 (2009).
\newblock \eprint{0812.3413}.
\newblock \doi{10.1103/PhysRevD.80.043510}

\bibitem{Serra:2008wc}
P.~Serra, A.~Cooray, Phys. Rev. \textbf{D77}, 107305 (2008).
\newblock \eprint{0801.3276}.
\newblock \doi{10.1103/PhysRevD.77.107305}

\bibitem{Hu:2001fa}
W.~Hu, Phys. Rev. \textbf{D64}, 083005 (2001).
\newblock \eprint{astro-ph/0105117}.
\newblock \doi{10.1103/PhysRevD.64.083005}

\bibitem{Kesden:2003cc}
M.H. Kesden, A.~Cooray, M.~Kamionkowski, Phys. Rev. \textbf{D67}, 123507
  (2003).
\newblock \eprint{astro-ph/0302536}.
\newblock \doi{10.1103/PhysRevD.67.123507}

\bibitem{Hirata:2002jy}
C.M. Hirata, U.~Seljak, Phys. Rev. \textbf{D67}, 043001 (2003).
\newblock \eprint{astro-ph/0209489}.
\newblock \doi{10.1103/PhysRevD.67.043001}

\bibitem{Hu:2001kj}
W.~Hu, T.~Okamoto, Astrophys. J. \textbf{574}, 566 (2002).
\newblock \eprint{astro-ph/0111606}.
\newblock \doi{10.1086/341110}

\bibitem{Okamoto:2003zw}
T.~Okamoto, W.~Hu, Phys. Rev. \textbf{D67}, 083002 (2003).
\newblock \eprint{astro-ph/0301031}.
\newblock \doi{10.1103/PhysRevD.67.083002}

\bibitem{Smith:2008an}
K.M. Smith, et~al., AIP Conference Proceedings \textbf{1141}, 121 (2008).
\newblock \eprint{0811.3916}

\bibitem{Seljak:1999zn}
U.~Seljak, M.~Zaldarriaga, Astrophys. J. \textbf{538}, 57 (2000).
\newblock \eprint{astro-ph/9907254}.
\newblock \doi{10.1086/309098}

\bibitem{Yoo:2008bf}
J.~Yoo, M.~Zaldarriaga, Phys. Rev. \textbf{D78}, 083002 (2008).
\newblock \eprint{0805.2155}.
\newblock \doi{10.1103/PhysRevD.78.083002}

\bibitem{Reichardt:2008ay}
C.L. Reichardt, et~al., Astrophys. J. \textbf{694}, 1200 (2009).
\newblock \eprint{0801.1491}.
\newblock \doi{10.1088/0004-637X/694/2/1200}

\bibitem{Hirata:2004rp}
C.M. Hirata, N.~Padmanabhan, U.~Seljak, D.~Schlegel, J.~Brinkmann, Phys. Rev.
  \textbf{D70}, 103501 (2004).
\newblock \eprint{astro-ph/0406004}.
\newblock \doi{10.1103/PhysRevD.70.103501}

\bibitem{Smith:2007rg}
K.M. Smith, O.~Zahn, O.~Dore, Phys. Rev. \textbf{D76}, 043510 (2007).
\newblock \eprint{0705.3980}.
\newblock \doi{10.1103/PhysRevD.76.043510}

\bibitem{Hirata:2008cb}
C.M. Hirata, S.~Ho, N.~Padmanabhan, U.~Seljak, N.A. Bahcall, Phys. Rev.
  \textbf{D78}, 043520 (2008).
\newblock \eprint{0801.0644}.
\newblock \doi{10.1103/PhysRevD.78.043520}

\bibitem{Perotto:2006rj}
L.~Perotto, J.~Lesgourgues, S.~Hannestad, H.~Tu, Y.Y.Y. Wong, JCAP
  \textbf{0610}, 013 (2006).
\newblock \eprint{astro-ph/0606227}

\bibitem{Das:2008es}
S.~Das, D.N. Spergel, Phys. Rev. \textbf{D79}, 043007 (2009).
\newblock \eprint{0809.4704}.
\newblock \doi{10.1103/PhysRevD.79.043007}

\end{thebibliography}

\end{document}